# Mechanical Properties of a Diamond Schwarzite: From Atomistic Models to 3D-Printed Structures

Levi C. Felix[1,2], Vladimir Gaál[1], Cristiano F. Woellner[3], Varlei Rodrigues[1] and Douglas S. Galvao[1,2]

[1] 'Gleb Wataghin' Institute of Physics, State University of Campinas, Campinas-SP, Brazil

[2] Center for Computational Engineering & Sciences, State University of Campinas, Campinas-SP, Brazil

[3] Physics Department, Federal University of Paraná, Curitiba-PR, Brazil

ABSTRACT

*Triply Periodic Minimal Surfaces (TPMS) possess locally minimized surface area under the constraint of periodic boundary conditions. Different families of surfaces were obtained with different topologies satisfying such conditions. Examples of such families include Primitive (P), Gyroid (G) and Diamond (D) surfaces. From a purely mathematical subject, TPMS have been recently found in materials science as optimal geometries for structural applications. Proposed by Mackay and Terrones in 1991, schwarzites are 3D crystalline porous carbon nanocrystals exhibiting the shape of TPMS. Although their complex topology poses serious limitations on their synthesis with conventional nanoscale fabrication methods, such as Chemical Vapour Deposition (CVD), TPMS can be fabricated by Additive Manufacturing (AM) techniques, such as 3D Printing. In this work, we used an optimized atomic model of a schwarzite structure from the D family (D8bal) to generate a surface mesh that was subsequently used for 3D-printing through Fused Deposition Modelling (FDM). This D schwarzite was 3D-printed with thermoplastic PolyLactic Acid (PLA) polymer filaments. Mechanical properties under uniaxial compression were investigated for both the atomic model and the 3D-printed one. Fully atomistic Molecular Dynamics (MD) simulations were also carried out to investigate the uniaxial compression behavior of the D8bal atomic model. Mechanical testings were performed on the 3D-printed schwarzite where the deformation mechanisms were found to be similar to those observed in MD simulations. These results are suggestive of a scale-independent mechanical behavior that is dominated by structural topology.*

## INTRODUCTION:

The advent of additive manufacturing paved the way for fabricating structures with very complex geometries, such as Triply-Periodic Minimal Surfaces (TPMS). They have been used as models for 3D-printed structures that were found to be very promising for structural applications [1-3]. At the atomic scale, some TPMS can be associated with stable negatively curved carbon materials known as schwarzites [4]. Recent works used optimized atomic structures to generate surface meshes which were 3D-printed for structural applications, such as load-bearing resistance [5] and hyper-velocity ballistic impacts [6]. There are many different schwarzite families [7], the Diamond (D) one (Figure 1) was investigated in this work. Both atomic and 3D-printed 4x4x4 supercell structures (Figure 1 and Table 1) were compressed beyond the linear regime in order to investigate their strength and deformation mechanisms.

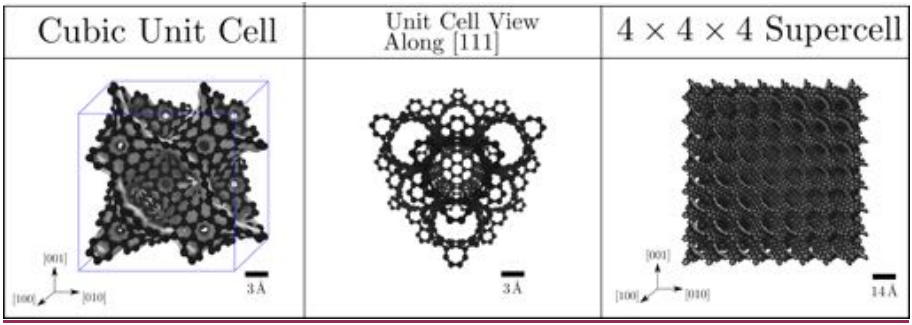

**Figure 1: D8bal schwarzite structure from the D family: cubic unit cell (left), its diagonal view (center) and 4x4x4 supercell (right).**

Table 1: Structural properties of D8bal structure. The number of atoms per cubic unit cell (n), in a 4x4x4 supercell (N), lattice parameter (a), structure length in one cubic direction of the atomic structures used in MD simulations ($l_{MD}$) and 3D-printed models ($l_{3D}$), and their corresponding specific masses ($\rho_{MD}, \rho_{3D}$).

| n | N | a [Å] | $l_{MD}$ [Å] | $\rho_{MD}$ [g/cm³] | $l_{3D}$ [mm] | $\rho_{3D}$ [g/cm³] |
|---|---|---|---|---|---|---|
| 768 | 49152 | 23.7 | 91.1 | 1.3 | 90.4 | 0.19 |

**METHODOLOGY:**

**Molecular Dynamics Simulations**

Fully atomistic Molecular Dynamics (MD) simulations were carried out using the Adaptive Interatomic Reactive Bond-Order (AIREBO) potential [8] as implemented by the Large-scale Atomic/Molecular Massively Parallel Simulator (LAMMPS) [9]. Reactive force fields, such as AIREBO, provides a better description of mechanical failure beyond the linear regime because it reproduces the breaking of bonds that occur on structural fractures. The uniaxial compression was simulated by applying a linearly moving reflecting wall at a rate of $10^{-5}$ $fs^{-1}$. In order to study the mechanical properties of solids, reactive MD simulations are an alternative to the finite element method, in which the latter strongly depends on the constitutive relation between stress and strain.

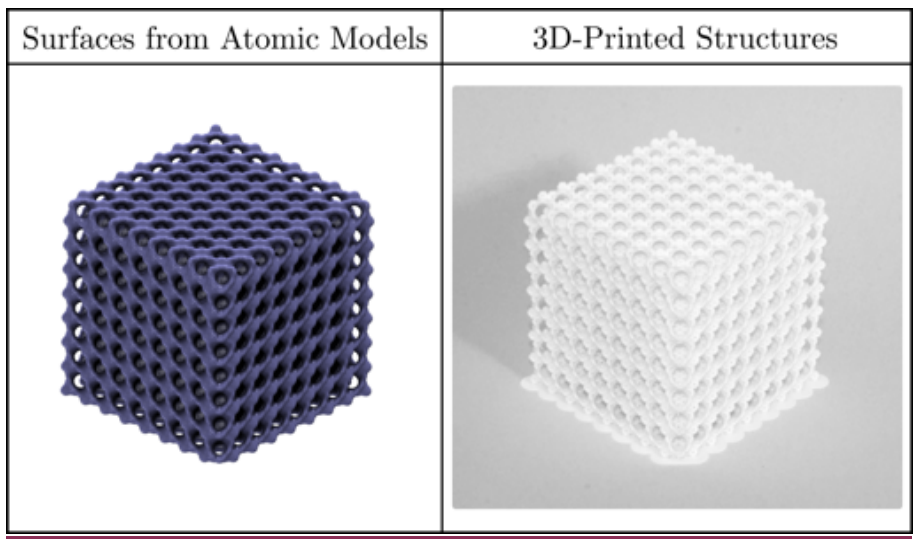

**Figure 2: Surface generated from the atomic model used in MD simulations (left) and 3D-printed polymer structure (right).**

### 3D Printing

Atomic structures from MD simulations were used as input to fabricate 3D printed schwarzites. The parts were 3D printed of PolyLactic Acid (PLA) using a homemade CoreXY Fused Deposition Modelling 3D printer. It uses a commercial hot nozzle of 400 μm in diameter to extrude PLA polymer filaments of 1.75 mm in diameter to print layers of 200 μm. The printing heads are moved along the XY plane by two stepper motors following the H-frame type XY-positioning system [10,11] and the hotbed is moved along the z-axis by another couple of stepper motors. The printer control is based on an open-source Arduino microcontroller board Mega 2560 interfaced with a commercial RepRap Arduino Mega Pololu Shield (RAMPS).

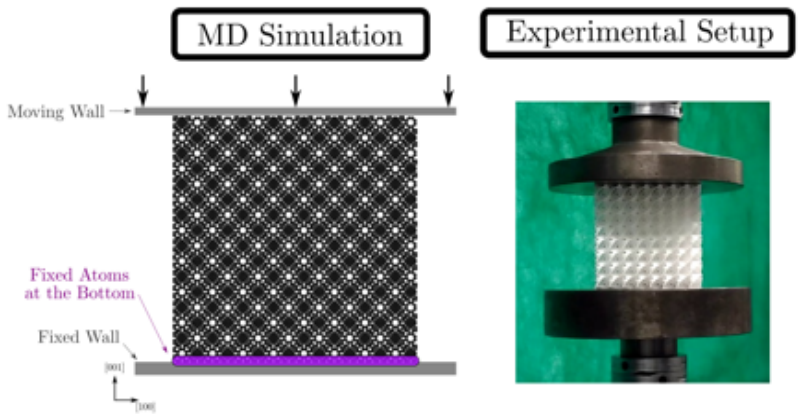

**Figure 3:** MD simulation setup (left) showing a linearly moving reflecting wall used to compress the structures while the bottom is kept fixed. Mechanical testing setup (right).

**Mechanical Testing**

To investigate the mechanical behavior of the structure the 3D-printed samples were compressed on an MTS 793 testing machine. This testing machine consists of a fixed table and a hydraulic piston that is set to compress the sample at a constant rate of 5 mm/min up to 80% strain. The hydraulic piston has a load cell to record the force during the test with a 200kN limit.

To record the transverse dimensions changes during the compression tests a Nikon D7100 camera was used with 1280 pixels resolution and 23 frames/s pointed perpendicular to one side of the sample.

A Python script was used to get frames from the recorded video at constant time intervals. The intensity profile for each frame was plotted along the [100] and [001] axis. As the sample was white and the background green the channel blue of the frame provides a white sample and black background. Thus, in the intensity profile, the high values are attributed to the sample while the low values to the background. Then, by analyzing the intensity profile it was possible to see that the slopes of intensity were the sample borders. Measuring the distance between the slopes (in pixels) and applying a scale factor (pixels per millimeter taken from the first frame) it was possible to measure the face dimensions (in millimeters) and consequently their change values.

**RESULTS AND DISCUSSION:**

In Figure 4 we present the results for the mechanical response under uniaxial compression for both atomic and 3D-printed schwarzites. D8bal accommodates considerable plastic deformation. Figure 5 shows how the D8bal deforms accordingly to MD simulations and mechanical testings on 3D-printed structures. Interestingly, two very different length scales show similar bending-dominated deformation under compression. A layer-by-layer-like deformation (both structures start to deform from the top) is observed both on MD simulations and 3D-printed structures. The Young's modulus values for the atomic and the 3D-print models are 108 GPa and 88 MPa, respectively.

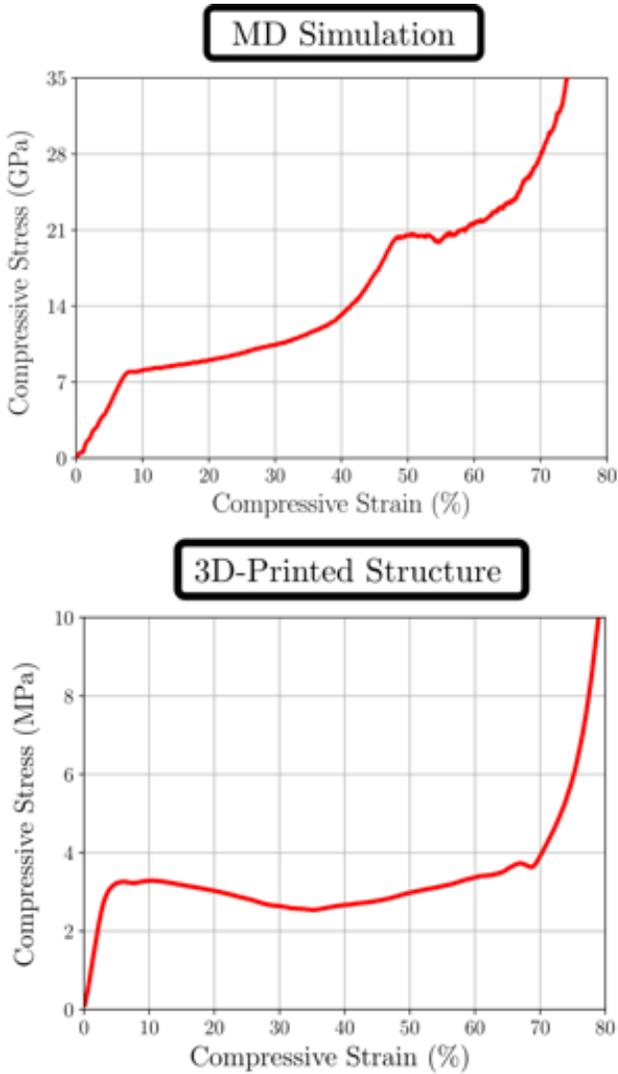

Figure 4: Stress-strain curve from MD simulations (top) and mechanical testing on 3D-printed structures (bottom).

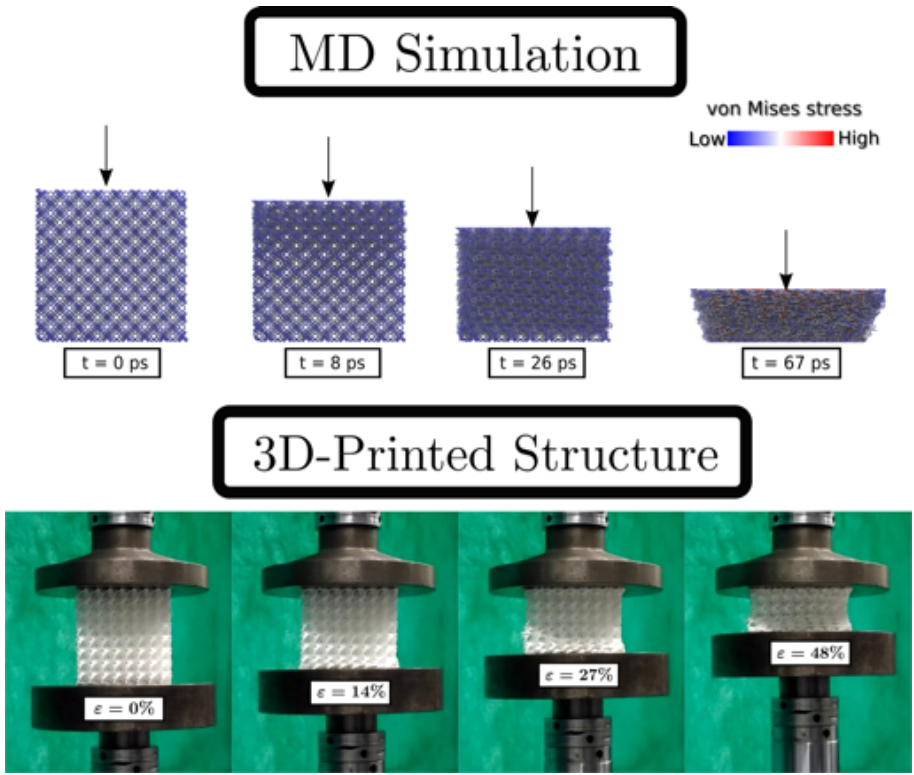

**Figure 5:** Snapshots from the MD simulations (top) and mechanical testing on 3D-printed structures (bottom), respectively. The elapsed simulation time and strain percentage are correspondingly indicated.

## CONCLUSIONS:

We investigated the mechanical properties (deformations under compressive strains) of the D8bal diamond schwarzite. We carried out fully atomistic reactive molecular dynamics simulations. The optimized atomic models were used to generate macro models that were 3D printed. Surprisingly, some of the observed mechanical properties are scale-independent, such as layer-by-layer deformations, which suggest that the topology features dominate the deformation mechanisms.